\documentclass[preprint2]{aastex}
\usepackage{emulateapj5}
\usepackage{onecolfloat5}
\slugcomment{Submitted to the Astrophysical Journal Letters, 11 Feb 2002; 
	accepted 19 March 2002.}

\newcommand{\etal}{{\it et~al.\/}\ }
\newcommand{\minusone}{$^{-1}$}

\shorttitle{Discovery of Variability in OH Megamasers}
\shortauthors{Darling \& Giovanelli}
 
\begin{document}
\twocolumn[
\title{The Discovery of Time Variability in OH Megamasers}
\author{Jeremy Darling \& Riccardo Giovanelli}
\affil{Department of Astronomy and National Astronomy and Ionosphere Center, 
	Cornell University, 524 Space Sciences Building, Ithaca,  NY  14853;
        darling@astro.cornell.edu; riccardo@astro.cornell.edu}
\begin{abstract}
We report the discovery of variability in the OH megamaser in IRAS
21272+2514 at $z=0.15$.  
This is the first OH megamaser (OHM) observed to vary in time.  The variation
is broadband, spanning 0.5--1.5 MHz or 100--300 km s\minusone\ in the 
rest frame, and strong, showing {\it rms} modulations of 10$\%$--15$\%$ while
the largest change in flux density between extremal epochs is 34$\%$.
Timescales sampled range from 39 to 821 days, and the characteristic 
modulation timescale is likely to be less than the shortest time baseline.  
The source
of modulation is currently ambiguous, although we favor interstellar 
scintillation.
Best estimates of the size scales of the 1667 MHz OH line features obtained
from refractive interstellar scintillation models constrain variable features
to be smaller than 2 pc, and quiescent features to be larger than 
a few pc.  Compact masers account for roughly 27$\%$--58$\%$ of the total
OH emission in this source.  Accelerations in spectral lines are generally
constrained to be less than 3 km s\minusone\ yr\minusone, but acceleration
of this magnitude is suggested in one of the quiescent spectral components
and merits further study.
\end{abstract}
]

\section{Introduction}
The OH megamaser (OHM) in IRAS 21272+2514 was discovered in 
August 1999 and reobserved 
in October 1999 as part of the Arecibo\footnote{The Arecibo
Observatory is part of the National Astronomy and Ionosphere Center, which 
is operated by Cornell University under a cooperative agreement with the
National Science Foundation.} OHM survey (Darling \& Giovanelli
2000, 2001, 2002a).  Spectra of this strong megamaser spanning 
85 days showed a remarkable change in some, but not all, spectral features.  
This is the first documented detection 
of variability in an OHM, and it suggests that many OHMs may exhibit 
variability on time scales of days to months.  Observations of variability 
can segregate maser emission into compact (variable) and extended (quiescent)
emission in a similar and complementary manner to very long baseline 
interferometry (VLBI) 
observations which show that roughly two-thirds of OHM emission is compact
\citep{lon98,dia99,pih01}.

IRAS 21272+2514 has a heliocentric optical redshift 
$z=0.15102\pm0.00012$, while the bulk of the multi-component OH emission 
is blueshifted
with respect to the optical, spanning 849 km s\minusone\ in the rest frame
at 10$\%$ of peak flux density and centered on $z_{OH}=0.15021\pm0.00005$
(Darling \& Giovanelli 2000, 2002b).  The integrated OH luminosity is among 
the highest known at $L_{OH} = 4.57 \times 10^3 L_\odot,$\footnote{We assume 
$H_\circ = 75$ km s\minusone\ Mpc\minusone, 
$\Omega_M = 0.3$, and $\Omega_\Lambda$ = 0.7.}
and the far-IR (FIR) luminosity is constrained to be in the range 
$L_{FIR} = (4.9$--$7.8)\times10^{11} L_\odot$ (the 100 $\mu$m flux is undetected
by IRAS).  The 1.4 GHz flux density is modest, at $4.4\pm0.5$ mJy
\citep{con98}, and IRAS 21272+2514 has some IR excess but is 
broadly consistent with the radio-FIR relationship for star forming galaxies
(Darling \& Giovanelli 2002a).  
This object has a Seyfert 2 nucleus (Darling \& Giovanelli 2002b).

In this letter, we present follow-up observations of IRAS 21272+2514 
designed to confirm the observed variability (\S \ref{obs}), we discuss the
modulation present in the OH spectral line components (\S \ref{modulation}),
we set limits on the size scales of all observed maser components
(\S \ref{sizes}),
we constrain 
accelerations in the OH spectral lines (\S \ref{acceleration}), 
and we discuss the implications of variability for 
understanding OHMs and the environments which produce them 
(\S \ref{conclusions}).

\begin{figure}[!ht]
\epsscale{0.9}
\plotone{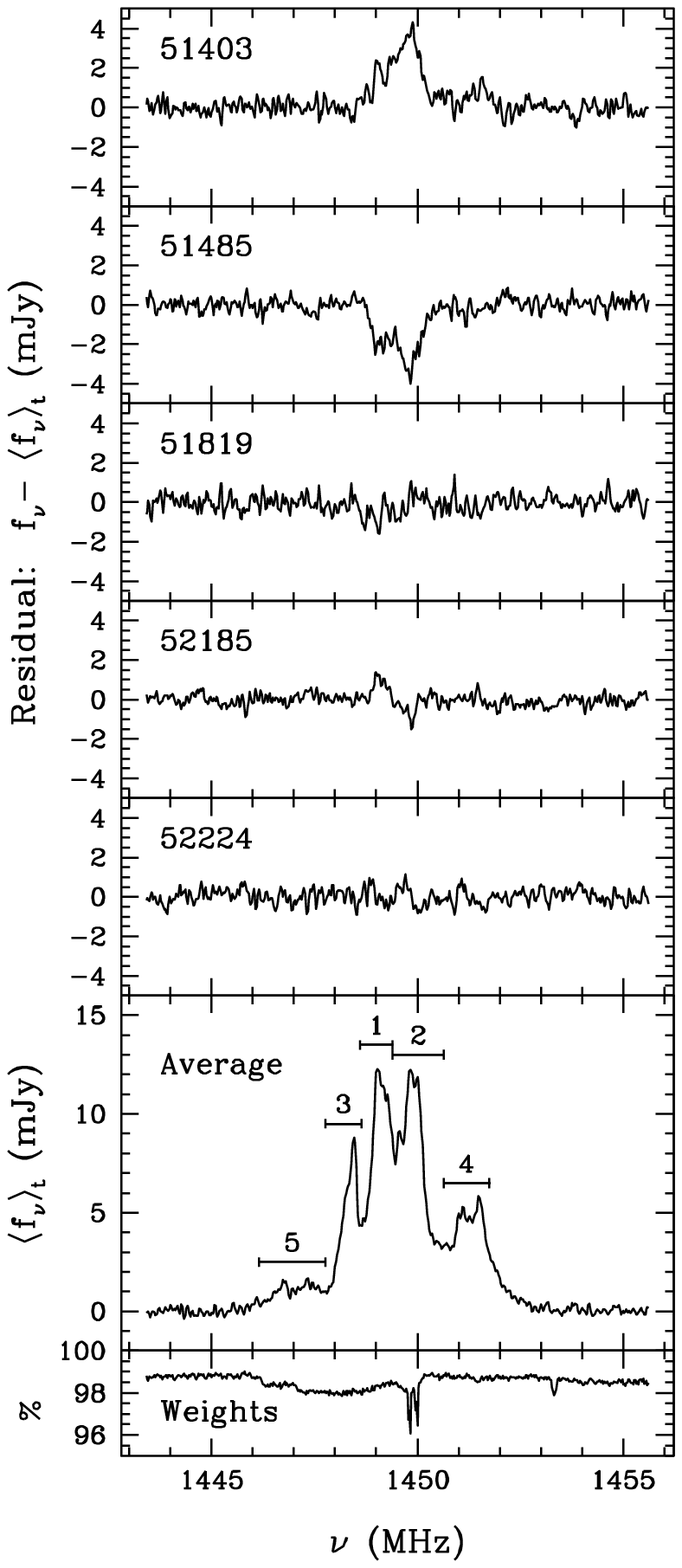}
\caption{Residual spectra of IRAS 21272+2514 in the five consolidated
epochs, each labeled by the MJD (Table \ref{journal}), and the 
average spectrum with emission line components labeled.  
The spectra bracket an 821 day interval.  The residuals show 
strong variability in this source, but only in components 1 and 2
of the emission line spectrum.  Spectra have been scaled to account for
broadband changes in sensitivity and calibration as described in \S 
\ref{modulation}.  The weights spectrum indicates the percentage of 
data records used to construct each channel in the average spectrum and 
indicates channels where RFI was flagged and removed.
\label{showdiff}}
\end{figure}

\begin{figure}[!ht]
\epsscale{0.98}
\plotone{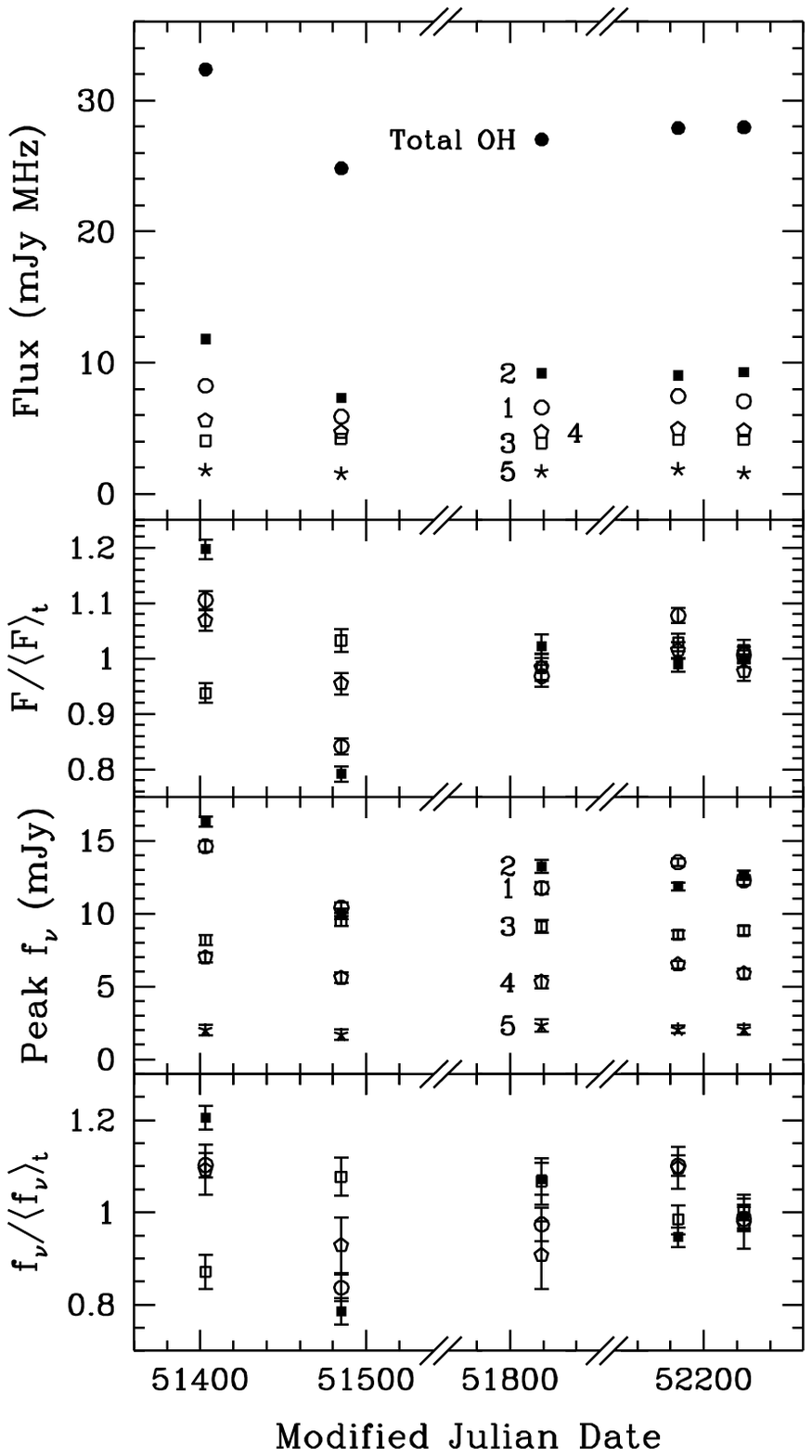}
\caption{Integrated line flux and peak flux density measurements of the OH 
emission line components by epoch.  
Components are identified in Figure \ref{showdiff}.
The abscissa has been shortened to facilitate comparison of epochs.  
Component 5, which has low signal to noise and large error bars,
is not shown in the residual plots.
\label{flux}}
\end{figure}

\section{Observations and Data Reduction}\label{obs}
Observations were performed at Arecibo with the L-narrow and L-wide receivers
as indicated in Table \ref{journal}, nodding on- and off-source for 4 minutes
each.  We record spectra in 1-sec intervals in two circular polarizations to 
facilitate radio frequency 
interference (RFI) excision as described in \cite{dar01}.  A weights spectrum
is produced for each epoch which shows the fraction of RFI-free records used
to construct each channel in a spectrum.  The weights spectrum spanning
all epochs used to construct the average spectrum shown in Figure 
\ref{showdiff} indicates that while some RFI is present in 1$\%$--4$\%$ of the
records, it is not the source of the observed spectral variation.  
Final spectra are averages of two circular polarizations which do not  
significantly differ except in channels with significant RFI.  

Observations have been consolidated into five epochs to enhance the spectral
signal to noise as indicated in Table \ref{journal}.  Effective modified
Julian dates (MJD) are time-weighted averages of observing sessions within
each epoch.  Although the separation of observations within epochs spans
1--8 days, no variability is observed within epochs.  Note that intraepoch 
variability is poorly constrained due to low signal-to-noise ratio.  

\begin{deluxetable}{ccccccc}
\tabletypesize{\footnotesize}
\tablecaption{Journal of Observations and Epoch Consolidation Scheme
\label{journal}}
\tablewidth{0pt}
\tablehead{
\colhead{Date} & \colhead{Center UT}  & \colhead{Receiver\tablenotemark{a}} 
& \colhead{$t_{on}$} & \colhead{MJD} & \colhead{Effective MJD} & \colhead{$t_{tot}$}\\
\colhead{(UT)} & \colhead{(hr)} & \colhead{} & \colhead{(minutes)} 
& \colhead{} & \colhead{}  & \colhead{(minutes)}  
}
\startdata
10 Aug 1999 & 04.60 & L-wide   & 20 &	51400.1917 \\
18 Aug 1999 & 04.83 & L-wide   & 12 &	51408.2012 & 51403.1953 & 32\\
25 Oct 1999 & 23.53 & L-wide   &\phn4&	51477.9804 \\
03 Nov 1999 & 22.39 & L-wide   & 28 &	51485.9329 & 51484.9388 & 32\\
29 Sep 2000 & 01.10 & L-narrow & 12 &	51816.0458 \\
05 Oct 2000 & 00.56 & L-narrow & 12 &	51822.0233 & 51819.0345 & 24\\
02 Oct 2001 & 01.33 & L-narrow & 16 &	52184.0554 \\
03 Oct 2001 & 01.33 & L-narrow & 16 & 	52185.0554 & 52184.5554 & 32\\
09 Nov 2001 & 23.34 & L-narrow &\phn8&	52222.9725 \\
10 Nov 2001 & 23.29 & L-narrow & 12 &	52223.9671 \\
12 Nov 2001 & 23.18 & L-narrow &\phn8&	52225.9658 & 52224.2540 & 28
\enddata
\relax\\[-2.5ex]
\tablenotetext{a}{There are two L-band receivers in the Gregorian system at 
Arecibo: the L-wide receiver (1.12--1.73 GHz), and the L-narrow receiver 
(1.28--1.50 GHz).}
\end{deluxetable}

Figure \ref{showdiff} shows the residual spectra (each epoch minus the average
spectrum) from the five consolidated 
epochs, the average spectrum over all time, and the weights spectrum of the
average.  The main emission components are labeled in the average spectrum.
The {\it rms} noise is 0.17 mJy in the average spectrum and 
typically 0.35 mJy in the residuals.  The residual spectra show
significant modulation only in emission line components 1 and 2.  
Spectra have been scaled to account for
broadband changes in sensitivity and calibration as described in \S 
\ref{modulation}.

\section{Spectral Line Modulation}\label{modulation}

Comparing measurements from different observing epochs can potentially 
introduce significant spurious time-dependent signals.  The observations 
presented here bracket significant changes in the performance of the 
telescope,
including a gross and fine refiguring of the primary surface, implementation 
of active control of the pointing with tie-downs anchoring the platform, 
refined pointing models, and general fine-tuning of the system.  
Observations were also performed with two different feeds (see
Table \ref{journal}).  The solution to the significant changes between 
observing epochs lies in accurate calibration of the gain for each 
observing session, in the broadband nature of the changes in telescope
performance, and 
in the complicated, multi-component spectrum of IRAS 21272+2514.  
Measuring {\it relative} changes between spectral features or changes 
of a given line as a fraction of the whole spectrum can resolve broadband
calibration issues.  Sources of spurious narrow-band variability (RFI) 
have been identified by their short timescales and excised.  

Measurements of relative variations between spectral features or with respect
to the total OH spectrum is complicated by the possibility that all spectral
features may vary, and that the variable components make up a significant 
fraction of the total OH flux.  We approach this problem iteratively, first
identifying both the variable and the apparently quiescent features, then 
comparing the fluxes of quiescent features to the total quiescent flux over
time to constrain the modulation actually present in each feature.  Once 
stable spectral features have been identified, the variable features are
measured with respect to the total quiescent flux to eliminate broadband
instrumental variability between epochs.  

Figure \ref{flux} shows for each spectral component labeled in Figure 
\ref{showdiff} the integrated flux density, the peak flux 
density, and the residual change in both scaled by the integrated 
spectrum minus spectral components 1 and 2.  The average peak and integrated
flux density of each spectral component is listed in columns (2) and (3) of 
Table \ref{vartab}.  
We find that components 1 and 2 show significant modulation of 
$10.4\%\pm1.6\%$ and $14.4\%\pm1.6\%$, respectively, while the remaining
components do not show significant modulation (col.\ [4] of Table 
\ref{vartab}).
The source modulation index $m_s$ is the {\it rms} fluctuation of the 
integrated line flux density.

A useful diagnostic of the temporal structure of variability is the 
structure function:  $D(\tau) = \langle[F(t) - F(t+\tau)]^2\rangle_t$.  
The structure functions of lines 3 and 4 are consistent with zero for all
10 time baselines, whereas lines 1 and 2 show large and highly 
variable values which indicate characteristic
timescales less than the shortest sampling interval of 39 days.

\section{Maser Size Scales}\label{sizes}
If the observed modulation is intrinsic to the source, then
the usual light travel time constraint applies to the amplified continuum 
source rather than the masing gas if the masing is unsaturated 
(unless the variability is due to motions
in the gas itself).  Variation is observed on the shortest
time baseline of 39 days, indicating that the variable source is no larger
than 0.03 pc or 6.7 $\mu$ac.  If the masing region is of roughly the same 
size, it would have a brightness temperature of $T\approx6\times10^{13}$ K.
Such a small source would certainly scintillate.  Hence, 
the observed variability is at least partly due to interstellar scintillation
(ISS), and we choose the weaker constraint on the angular sizes of the
variable masing regions by adopting ISS to interpret the observations.

\citet{wal98} provides rough predictions of the scintillation properties of 
extragalactic sources based on the interstellar medium model of \citet{tay93}
which we apply to the observed variation in the OH line components of 
IRAS 21272+2514.  The OH line radiation 
must pass through two scintillating screens, but the dominant influence
will be from the electron screen in our own galaxy.  
The galactic coordinates of 21272+2514,
$(l,b) = (75.7^\circ,-18.4^\circ)$, indicate a transition frequency
from strong to weak scattering at $\nu_\circ = 15 \pm 1$ GHz 
and a first Fresnel zone at
$\nu_\circ$ of size $\theta_{F\circ} = 2\ \mu$as.  At $\nu=1.45$ GHz,
the scattering should be strong refractive ISS (RISS).  For a point
source, \citet{wal98} predicts a modulation index of $m_p=27\%$,
a scattering disk 0.34 mas in size, and a scintillation timescale
of 10 days.  We scale the point source modulation index $m_p$ to match 
the observed modulation $m_s$ in each source component to obtain rough
estimates of source sizes and scintillation timescales from the
scaling relations $\theta_s/\theta_p = (m_p/m_s)^{6/7}$ and 
$\tau_s/\tau_p = \theta_s/\theta_p$.  Estimates of the angular
sizes, RISS timescales, and physical sizes of the OH line components are
listed in columns (5)--(7) of Table \ref{vartab}.  
The two components which show strong modulation
have submilliarcsecond angular sizes, are less than 2 pc in size, and
are expected to show a characteristic RISS timescale of 17--22 days.  This
is in good agreement with the observations and our inability to observe 
changes on short time baselines of 1--8 days.  This treatment assumes
that all of the emission within a variable spectral component originates from
the same compact region.  It is also possible that part of the emission
is compact and part is extended, so the scintillation predictions are
actually upper limits on the source size and the RISS timescales may 
be as low as 10 days.  If the compact variable features are smaller
than the scattering disk, then they are expected to show the full point 
source modulation of 27$\%$ and
must account for at least 39$\%$ and 53$\%$ of the emission seen in components
1 and 2, respectively.  In other words, no more than roughly half of the 
emission in components 1 and 2 can be extended.  
These constraints on the fraction of compact 
emission present in OHMs are consistent with VLBI 
observations of nearby 
OHMs by \citet{dia99} and others who find that roughly two-thirds of the total 
OH emission is compact.  

Components 3--5, which do
not show significant modulation, are constrained to be larger than 1
mas or a few parsecs.  RISS timescales in these components are 
constrained by their modulation indices to be larger than 28--51 days.
Note that although observations span longer time baselines, the observed
limits on modulation provide the stronger constraint --- modulation 
may be present in components 3--5 on timescales of a few months, but 
the observations lack the sensitivity to detect variations of less than about
4$\%$.  The fraction of compact emission in components 3, 4, and 5 is 
less than 15$\%$, 16$\%$, and 30$\%$, respectively.

\begin{deluxetable}{cccrccccc}
\tabletypesize{\footnotesize}
\tablecaption{OH Line Modulation, Scintillation Predictions, and 
Acceleration in IRAS 21272+2514\label{vartab}}
\tablewidth{0pt}
\tablehead{
\colhead{} & \colhead{} & \colhead{} & \colhead{} & 
\multicolumn{3}{c}{Scintillation Predictions}\\ \cline{5-7}
\colhead{Feature} & \colhead{$\langle f_\nu\rangle_t$\tablenotemark{a}} 
& \colhead{$\langle F\rangle_t$\tablenotemark{b}} 
& \colhead{$m_s$} 
& \colhead{$\theta_s$} 
& \colhead{$\tau_s$} 
& \colhead{$D_s$} 
& \colhead{$\langle\nu_{pk}\rangle_t$} & \colhead{$a$} \\
\colhead{} & \colhead{(mJy)} & \colhead{(mJy MHz)}  & \colhead{($\%$)} 
& \colhead{(mas)}  
& \colhead{(days)}  
& \colhead{(pc)}  
& \colhead{(MHz)}  
& \colhead{(km s\minusone\ yr\minusone)}\\
\colhead{(1)}& \colhead{(2)}& \colhead{(3)}& \colhead{(4)}
& \colhead{(5)}& \colhead{(6)}& \colhead{(7)}& \colhead{(8)}
& \colhead{(9)} 
}
\startdata
1 & 12.3 &  7.06 & $10.4\pm1.6$ & $<0.76$ & 10--22 & $<1.9$ & $1449.099\pm0.031$ &\phs$ 3.4\pm3.6$ (5)\\
2 & 12.2 &  9.27 & $14.4\pm1.6$ & $<0.58$ & 10--17 & $<1.4$ & $1449.901\pm0.029$ &    $-5.5\pm2.4$ (4) \\
3 &\phn8.8& 4.13 & $ 4.0\pm1.9$ & $>1.74$ & $>51$ & $>4.2$ & $1448.523\pm0.013$ &    $-2.6\pm0.8$ (5) \\
4 &\phn5.3& 4.93 & $ 4.4\pm1.9$ & $>1.60$ & $>47$ & $>3.9$ & $1451.560\pm0.017$ &    $-2.6\pm1.6$ (5) \\
5 &\phn1.7& 1.59 & $ 8.0\pm4.3$ & $>0.96$ & $>28$ & $>2.3$ & \nodata & \nodata \\
\enddata
\relax\\[-2.5ex]
\tablenotetext{a}{Uncertainties in the peak flux densities are 
$\pm0.2$ mJy, excluding possible systematic calibration errors.}
\tablenotetext{b}{Uncertainties in the integrated line flux densities are 
$\pm0.03$ mJy MHz, excluding possible systematic calibration errors.}
\end{deluxetable}

\section{Acceleration}\label{acceleration}

Like water masers associated with AGN, one might expect OH megamasers to 
exhibit accelerations in emission line features.  The kinematical structure
of major mergers is certainly complicated and dynamic, and it is 
possible that accelerating masing gas will be observable over the timescales
of this study.  Water maser features in the best-studied case, NGC 4258,
show accelerations of 9.3 km s\minusone\ yr\minusone\ along the line of sight
\citep{her99}.
Accelerations of this magnitude in OHM line features can be detected 
by this study if they exist.  One might expect that since OHM emission
profiles are thought to be an ensemble of many masing regions, then 
accelerations will be averaged out by the summation.  This is especially 
true if OHMs are produced in highly turbulent or chaotic accumulations of
molecular gas.  The wide range of line profiles suggests that this may be
the case, although there are a number of narrow, physically compact 
emission lines in the sample.  It might be surprising if the lines
showing intensity fluctuations do not show significant accelerations.

Overall, frequency/velocity residuals are consistent with the uncertainties
in the line peaks of $\pm1$ channel, corresponding to 24 kHz in the observer 
frame.  The mean frequency of the peak of each spectral component is listed 
in column (8) of Table \ref{vartab}.  The associated uncertainties are 
generally consistent with $\pm24/\sqrt{N}$ kHz where $N$ is the number of 
epochs used in the mean ($N$ is listed in parentheses at the end of col.\ [9]
of Table \ref{vartab}).  Acceleration can be measured from the slope of
a linear fit to the time series of line peak frequencies and expressed
in the rest frame of the source by including a factor $(1+z)^2$ to account
for time dilation and the narrowing of frequency intervals with redshift.  
The minimum detectable acceleration observable in an OHM line feature
is set by the uncertainty of the
best-fit slope obtained from $N$ observations with velocity resolution 
$\pm4.4$ km s\minusone (observer frame):
\begin{equation}
  \sigma_a = {\sigma_v (1+z) \over  \sigma_t\sqrt{N-1}} = 
	(4.4 \mbox{ km s\minusone\ yr\minusone}) 
		{(1+z)^2\over\sigma_t(\mbox{yr})\sqrt{N-1}}
\end{equation}
where $\sigma_t^2$ is the variance of the sampling in time and $\sigma_v$ is
the uncertainty of $\pm1$ channel in identifying the peak of a line.  
For 21272+2514, with five epochs sampled, $\sigma_a = 2.8$ km s\minusone\  
yr\minusone.  This sets a rough lower limit on the line accelerations which 
may be measured from the available data.  

Table \ref{vartab} lists the rest frame line accelerations for each 
component from fits to the time series of line peaks.
Uncertainties are obtained from the fits and $N$ is listed in parentheses.
Except component 2 where the peak could not be identified in one epoch,
accelerations are obtained from fits to all data points.  
Line accelerations are consistent with zero except for line 3 which shows
a 3.3 $\sigma$ deviation from zero.  The acceleration of $-2.6\pm0.8$ 
km s\minusone\ yr\minusone\ for component 3 is very close to the expected
lower detection limit of 2.8 km s\minusone\ yr\minusone.
Confirmation of this acceleration requires re-observation, which will provide
a longer time baseline.  It is worth mention that the peaks of line 3
show a steady decrease in frequency over time with little scatter and no
outliers.  It is interesting that this line also shows no significant 
modulation in amplitude.

\section{Conclusions}\label{conclusions}
This brief study of IRAS 21272+2514 demonstrates the powerful 
constraints that variability studies can place on the emission regions of OHMs,
including the segregation of different spectral components into different
physical size scales.  Better sampling of intermediate time scales can
provide even stronger constraints on the source properties including the
physical environments responsible for OHMs.  There is mounting evidence
that OHMs are composed of both diffuse and compact emission regions with 
different saturation states \citep{lon98,dia99}.  This study obtains 
constraints on both the size and proportion of compact emission regions:
compact features must be smaller than about 2 pc and account for 27$\%$--58$\%$
of the total OH emission.  To account for such high surface
brightness emission we suggest that either shocks are involved in the
pumping of OH or that these compact regions represent fortuitous 
geometries which provide enhanced masing path lengths as suggested by 
\citet{pih01}.  Accelerations of the OH lines of IRAS 21272+2514 
are generally constrained to be less than 3 km s\minusone\ yr\minusone,
although further observations are required to resolve suggestions
of acceleration in one component.

\acknowledgements
Many thanks indeed to B. Catinella for obtaining the spectra of IRAS
21272+2514 during the MJD 52185 and 52224 epochs and to M. Walker for 
supplemental scintillation calculations.  It is a pleasure to thank the 
staff of the Arecibo Observatory for excellent support.  
This work has received support from STScI grant 8373 and NSF grant 
AST 00-98526.

\end{document}